# FORECASTING PANDEMIC TAX REVENUES IN A SMALL, OPEN ECONOMY: THE CASE OF BULGARIA

**Fabio Ashtar Telarico**
Centre for Southeastern European Studies, Graz, Austria
fabio.telarico@edu.uni-graz.at, fatelarico@gmail.com

*Abstract: Tax analysis and forecasting of revenues are of paramount importance to ensure fiscal policy's viability and sustainability. However, the measures taken to contain the spread of the recent pandemic pose an unprecedented challenge to established models and approaches. This paper proposes a model to forecast tax revenues in Bulgaria for the fiscal years 2020–2022 built in accordance with the International Monetary Fund's recommendations on a dataset covering the period between 1995 and 2019. The study further discusses the actual trustworthiness of official Bulgarian forecasts, contrasting those figures with the model previously estimated. This study's quantitative results both confirm the pandemic's assumed negative impact on tax revenues and prove that econometrics can be tweaked to produce consistent revenue forecasts even in the relatively-unexplored case of Bulgaria offering new insights to policymakers and advocates.*

*Key words: Forecast, taxation, VAT, personal income tax, Bulgaria*

## 1. INTRODUCTION

### 1.1. How the pandemic is changing forecasts

Over the last year or so, official and academic forecasts' inability to correctly predict at least the general trend behind upcoming developments has eroded confidence. So much that the Fiscal Affairs Department (FDA) of the International Monetary Fund (IMF) warned that the most often employed techniques to forecast tax revenue will not suffice to grasp the major decline in tax revenue that the pandemic will cause in most countries. (IMF FAD 2020, 1).

Similar issues have already puzzled the economic-policy establishment for a while now. Hence, new fiscal-forecasting models can take advantage of the important steps towards new models (Demertzis and Viegi 2008; Brubakk and Sveen 2009; Alessi et al. 2014) in the area of monetary-policy forecasting (Demertzis and Viegi 2020; Demertzis and Dominguez-Jimenez 2020). Yet, this vast knowledge is not a solution *per se* as it does nothing to reduce the novelty of the ways in which the pandemic is – and will continue – affecting the economy. Moreover, looking across the fence cannot be the solution since tax-revenue forecasting manifests more than a few peculiarities. Hence, this paper tries to put in place a workable approach to tax revenue forecasting by combing the FAD's recommendations with insights from non-fiscal forecasting.

### 1.2 Background of the study

This paper focuses on Bulgaria, a small, open and relatively backward economy. Despite EU membership, the country coins its own currency: The *Bulgarian Lev* (BGN). The BGN is pegged to the Euro (€) at a fixed exchange rate (BGN 1.98 for €1) (NS-RB 2020).

In 2019, GDP was around €60bln (NSI 2020b) or €8,701 per-capita (NSI 2020a). Bulgaria is the poorest EU country in Purchasing Power Standards, standing at 53% of the EU average (EUROSTAT 2021a). Furthermore, the documented economy does not represent the entirety of Bulgarians' activities given that estimates put the grey economy around 21% of GDP (Ivanov 2021), with peaks for employment (Toteva 2021) and tobacco up to 50% (*Mediapool.bg*'s Editors 2020).

In the past decade (2009–2019), tax revenues grew on average by 2.45% on a yearly basis. These figures lag far behind the average 5.10% yearly GDP growth over the same period. Thus, the tax-to-GDP ratio has decreased steadily by more than 6.5% touching the bottom in 2015. Bulgaria's tax revenues are also underperforming in comparison to EU countries which are members of the OECD, but not of the Eurozone. There, in the 2010s, nominal tax revenues have risen as a share of GDP by an average 3.25% annuum.

The revenue disaggregation in TABLE 1 **Error! Bookmark not defined.** shows that there was a partial improvement in tax revenues across the board in 2019. However, this development was a carry-over of the GPD's sheer positive dynamics (NSI 2020b) rather than being driven by policy. Moreover, this pick-up brought relatively few funds and the fiscal multipliers are going to be smaller in any case (Karagyozova-Markova, Deyanov, and Iliev 2013, 30). Fiscal policy's ineffectiveness is partly due to the high degree of informality, which determines difficulties in tax collection (Rosser, Rosser, and Ahmed 2000, 160) as well as revenue mobilisation (Dellas et al. 2017).

## Table 1: The decay of Bulgaria's tax revenue was on a recovery trajectory in 2019, but still far from 2009 levels

Revenue-to-GDP ratio for the main taxes in Bulgaria in 2015-2019 compared to their 2009 level

| Taxes | 2015 | 2016 | 2017 | 2018 | 2019 |
|---|---|---|---|---|---|
| Customs duties and taxes | 67 | 69 | 70 | 79 | 98 |
| Tax on insurance premiums* | 167 | 173 | 155 | 178 | 196 |
| Excise duty | 98 | 108 | 107 | 108 | 112 |
| Value added tax | 83 | 89 | 85 | 93 | 104 |
| Personal income taxes | 120 | 122 | 120 | 131 | 151 |
| Taxes on dividends, liquidation shares and income of legal entities | 52 | 68 | 36 | 40 | 57 |
| Corporate tax | 63 | 69 | 73 | 83 | 93 |
| Minor taxes | 56 | 59 | 55 | 61 | 66 |

* 2009=100 for each indicators, except for the tax on insurance premiums had not been introduced yet (2011=100).

**TABLE 3** Source: *Law on the State Budget of the Republic of Bulgaria* for 2010; 2016; 2017; 2018; and 2019.

However, the crisis induced by anti-contagion measures is almost certain to cause significant tax revenue losses. With all probability, no recover will be in sight until at least 2021. Thus, policy choices apt to spur new growth need a brave mid-term perspective embracing the postcrisis *recovery* phase (Demertzis 2021). Thus, understaing tax revenues' dynamics is vital for Bulgarian policymakers and citizens.

### 1.3 Research's aims

The objective of this research is to forecast the revenues generated by the value added tax (VAT) and the personal income tax (PIT) for the fiscal years 2020–2022. In doing so, this paper is prescient of both the indication provided by the IMF's FAD and the ECB's new methodology.

The second section describes a *baseline* scenario ignoring the pandemic and in which policies do not change significantly. The technique known as Autoregressive Integrated Moving Average (ARIMA) underlies this forecast.

The third section dives into a more *realistic* scenario, acknowledging the pandemic's impact and endeavouring to incorporate the effects stemming from the policies adopted to contain it. Here, estimates admit a greater degree of arbitrariness and contestability because of the necessary "subjective adjustments" to account for events which "are not captured by the model" (IMF FAD 2020, 8) for the baseline scenario.

Reasoned estimates are offered for a series of parameters then fed into a multivariate regression model to forecast revenues. These two forecasts are then compared to one another and to official governmental forecasts from the past years. The models are verified by several error statistics (mainly *root mean squared error*, RMSE), the *Diebold-Mariano* (DM) test, and *Theil's U* statistic ($U_1$).

## 2. Data and methodology

### 2.1 Collected data

The time series used cover the years from 1995 to 2019. The two forecasted scenarios attempt to predict personal income tax revenue (PIT) and value added tax revenue (VAT) in 2020–2022. Where needed because of scarce data (e.g., regarding excises and customs duties), necessary approximations are explained in the text.

First, the ARIMA model for both time series is estimated following all indications offered by the main works in the literature (Tiao and Box 1981; Harvey and Todd 1983; Box et al. 2016). This approach has been employed productively in the estimation of tax revenue for emerging economies in general (Streimikiene et al. 2018) and transition economies in particular (Legeida and Sologoub 2003).

Whenever a third variable is supposed to effect revenue, causation and directionality are proved applying the Granger (G-causality) test (Granger 1969). This technique also finds wide applicability to the analysis of taxation (Heckelman 2000; Tosun and Abizadeh 2005).

Finally, models' robustness were evaluated through RMSE (Nau 2014; Streimikiene et al. 2018), DM tests (Diebold and Mariano 2002), and Theil's U (DoT Australia 2008).

### 2.2 Unit root test

The first step is to verify the data series' stationarity. There are various means to proves whether a set of data is stationary by looking for a unit root. The Augmented Dickey–Fuller (ADF) test used here is probably the most widely of such proofs in academia to date (Dickey and Fuller 1979; 1981; Streimikiene et al. 2018, 728).

$$\Delta y_t = u_t + \alpha + \beta_t + (\rho - 1)y_{t-1} + \sum_{i=1}^{k} \vartheta_i \Delta y_{t-i} \quad (1)$$

### 2.3 Multivariate linear regression

There are several ways to describe a multivariate linear regression. Most notably, one can choose between the scalar form and matrixes. Opting for the latter, such a model can be built in a few steps (Arminger, Clogg, and Sobel 1995, 97) and summarised in the following equation:

$$y_i = X_i \beta + E_i \quad for\ i \in \{1, 2, \cdots, n\} \quad (2)$$

Stationarity is a useful property in regard to meeting (or supposing) the three assumptions of multivariate regression are met (*Ibid.*, 98–99).

### 2.4 Autoregressive integrated moving average (ARIMA) model

One of the best techniques to forecast future values in time series is the ARIMA model. Its basis is a mixed process including an autoregressive (AR) and a moving average (MA) component. The AR element is such that "the current value of the process is expressed as a finite, linear aggregate of previous values of the process and a random shock" (Box et al. 2016, 8–9) A stationary AR process is by *mean reverting* (Kirchgässner, Wolters, and Hassler 2013, 44). By contrast, the MA component expresses each deviation from past values as "linearly dependent on a finite number $q$ of previous" values (Box et al. 2016, 9). Obviously, MA processes are not mean-reverting. Thus, ARMA process partially absorb exogenous shocks via the AR component while being permanently modified by them due to the MA one. The $ARMA(p, q)$ model can be generalised as an $ARIMA(p, q)$ model where *d* determines the number of times the process has to be integrated before becoming stationary.

### 2.5 Forecast errors

Various techniques are employed to assess the relevance of the ARIMA model's forecasting error in comparison to official forecasts and alternative models: RMSE, which is considered as the "most reliable" of such indicators (Streimikiene et al. 2018, 731), Theil's U test (DoT Australia 2008, 7ff), and the DM test.

### 2.6 Granger causality test

In addressing how the behavioural and policy changes induced by the pandemic affect tax revenues it is necessary to establish causality between variables. The technique adopted to do so is the G-causality test, which is based on ordinary least squares regression (Granger 1969).

## 3. Estimations and discussion

### 3.1 Stationarity

Data stationarity is essential to then employ the ARIMA model and other analogous techniques.

The ADF test shows that both VAT and PIT are non-stationary at level. The unit root can be removed by differencing once.

**3.2 Baseline forecast**

The ARIMA model is used for forecasting purposes by taking into account the entire time first difference of the dataset going from 1995 to 2019 for both VAT and PIT. The auto-correlation function (ACF) and the partial ACF shows the best-fitting model is an $ARIMA(1,1,1)$.

In this way, PIT revenues in 2020 would be forecasted at €1.92bln, with an average yearly growth of 7.87% in 2020–2022. Analogously, VAT revenues would be €5.78bln and average a 6.68% yearly growth.

**3.3 Policy scenario — Approximating reality**

The alternative forecasts need to account for the fact that people were physically prevented from leaving their homes, with unpredictable effects on their capability to earn an income. Moreover, being self-induced, the crisis poses novel interrogatives. True, some sectors are benefitting from the current situation. However, new jobs can supply for less than a third of total dismissals in advanced economies (Barrero, Bloom, and Davis 2020, 11) and the unequalness of this "K-shaped recovery" (Telarico 2021) is undeniable.

Overall, the impact on consumption and incomes is likely to be even bigger for less-developed countries (Estupinan and Sharma 2020) by the reduction of 'grey' activities which grant survival incomes to poorer households (Narula 2020). Moreover, young and female workers are more likely to be hit by the crisis (Blustein et al. 2020; Chakraborty 2020; ILO 2020).

*3.4 Value-added tax*

The VAT is the main source of tax revenue for Bulgaria's budget. Its regressive effects, due to poor households' higher propension to consume (Carroll and Kimball 1996; Carroll et al. 2017; Morozumi and Acosta Ormaechea 2019; Fisher et al. 2020), are well-known and studied (Tamaoka 1994, 60–69; Kato 2003, 3). However, the pandemic has shown how downward-flexible consumption can be (Brinca, Duarte, and Faria-e-Castro 2020) when an exogenous shock strikes at the same time both demand and supply across a number of sectors (Guerrieri et al. 2020). In forecasting Bulgaria's VAT revenues, it is difficult to estimate exactly how the collapse in foreigners' arrivals and internal tourism will affect VAT revenues — mainly because of a lack of data. Moreover, the government has intervened by lowering VAT rates on books, children's food, and diapers halfway into the fiscal year (Lex.bg's Editors 2020). The only data the National Statistical Institute (NSI) releases regarding added value (AV) are aggregated according to NACE criteria (EUROSTAT 2008, 57).

Thus, the following alternative takes as a starting point VAT revenues' elasticity to GDP and final consumption. Then, VAT revenues for 2009–2019 are estimated by feeding a multivariate-regression model with a summary of GDP (TABLE 3), final-consumption (EU Commission 2020, 175) and C-efficiency forecasts (Author's calculations based on the formula shown in Keen 2013, 427ff). The granger test proves causality between each of these independent variables and VAT revenues.

**Table 3: GDP forecasts for Bulgaria and the Euro Area in 2020-2022 were summarised and fed to a multivariate-regression model**

| (3.A) | Forecast | Year | Euro Area | Bulgaria |
|---|---|---|---|---|
| | Actual data | 2019 | 1% | |
| | Autumn 2020* | 2020 | −8% | −5% |
| | Autumn 2020* | 2021 | 4% | 3% |
| | Autumn 2020* | 2022 | 3% | 4% |
| | Spring 2020* | 2020 | −8% | |
| | Spring 2020* | 2021 | 6% | |

| (3.B) | Forecast | Year | Euro Area | (3.C) | Forecast | Year | BGR |
|---|---|---|---|---|---|---|---|
| | | 2019 | | | | 2019 | |
| | Mild† | 2020 | −6% | | Summary‡ | 2020 | −3% |
| | Mild† | 2021 | 7% | | Summary‡ | 2021 | 1% |
| | Mild† | 2022 | 2% | | Summary‡ | 2022 | 5% |
| | Severe† | 2020 | −13% | | | | |
| | Severe† | 2021 | 3% | | | | |
| | Severe† | 2022 | 4% | | | | |

**TABLE 2** Sources: *EUROSTAT (in EU Commission 2020, 172); † ECB (2020); ‡ Author's calculations.

On this basis, VAT revenues in Bulgaria could have fallen by as much as 17% in 2020. The recovery in 2021 and 2022 would be rather slow and unsatisfying (5% and 6% respectively), with revenues still at 96% of their 2019 level by the end of 2022. A reduction in compliance and collection efficiency – estimated through a drop in C-Efficiency – would mean a 10% fall in VAT revenues as a share of GDP. Against the background of growing government expenditures (EU Commission 2020, 175), a decline in VAT revenues would probably lead to budget disbalances. Thus, the upward trend on which Bulgaria's debt and deficit were already set should persist.

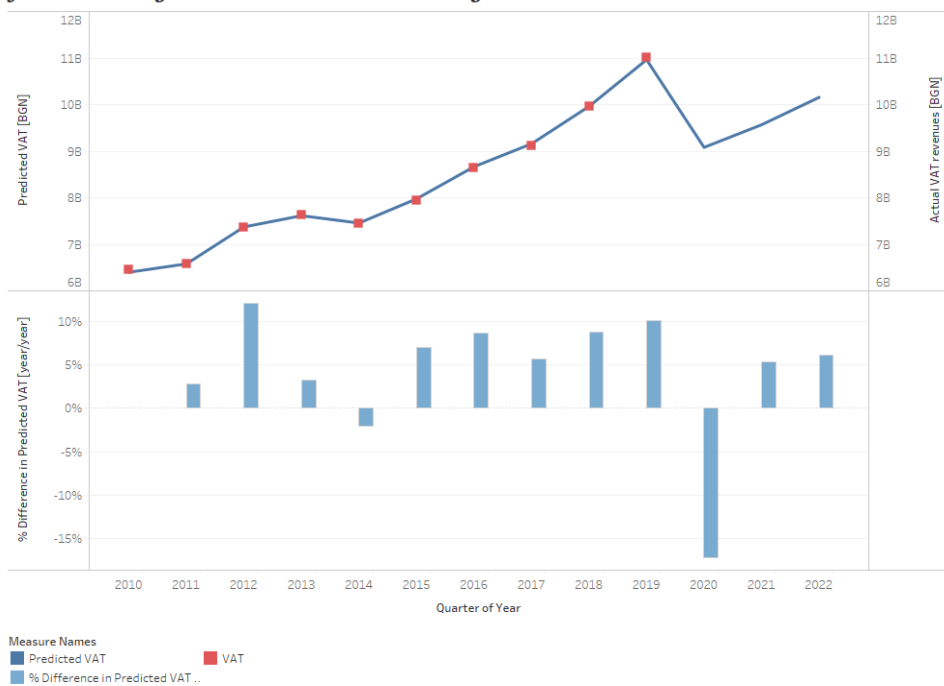

**Chart 2**: *According to the multivariate-regression model, VAT revenues are expected to fall drastically in 2020 and not to recover by 2022*

**Chart 1** Sources: Official data of the Bulgarian government, Author's calculations.

*3.5 Personal Income Tax*

PIT is the second largest source of tax revenue for the State's budget. Clearly, the relation between income and consumption does not always translate in more consumption; yet, falling incomes almost always cause the latter's contraction (Diacon and Maha 2015; for transition economies see: Kolasa 2012). In the case of Bulgaria, one needs to remember that about half of the households depend on employment income (47.2%) (EUROSTAT 2020b). Thus, the pandemic has likely impacted the livelihood of, at least, about half of Bulgarian households.

There are only a few datasets offering fine-grained data on the crisis's impact on various sorts of income. The only such data for Bulgaria are EUROSTAT's (2017) *flash estimates* on "income inequality and poverty". Together with structural data on Bulgarian households' income, pension and social benefits, these statistics represent the independent variables used to feed a multivariate-regression model. Total pensions (PEN), total social transfers (SOC) and total employment income (WAGE) were chosen as predictors because taken together, they account for over 99% of average disposable income (NSI 2020c).

It is relatively easy for policies to act on the factors determining PIT revenues. In fact, the Bulgarian government established a furlough and other wage-protection schemes (Milcheva 2020; Draganov 2020) reducing the crisis's negative effect on employment incomes by about a third (EUROSTAT 2020c). Besides that, the government has also accelerated the tempo of pensions' increase (NOI 2020a; 2020b; see also Darik news' Editors 2020). Given PEN's and SOC's clear G-causality on PIT revenues, the stimuli provided through these two channels are sure to generate an increase in the latter. Moreover, it is worth noting that these two sources account for more than half of the average Bulgarian household's disposable income (NSI 2020c).

Lastly, a few considerations on the forecasts fed into the model and shown in TABLE (below). EUROSTAT (2020c; 2020a) foresees Bulgarians' wages falling 3.12% in 2020. Given that wages are historically very elastic to growth, forecasted GDP values are employed to estimate WAGE's growth: 1.43% in 2021 and 4.92% in 2022. Using the recent official data (NOI 2021), pensions' growth is posited to be 15.79% in 2020. Assuming that in 2021 there will be a reindexing of pensions (Trud's Editors 2020) or another marked increase (Nikolova 2020; Blitz's Editors 2021), the figure is put at 18%.

**Table 4:** *Summary of forecasts for the three determinants of PIT revenues in Bulgaria*

|      | Δ SOC | Δ WAGE | Δ PEN  | Δ GDP  |
|------|-------|--------|--------|--------|
| 2010 |       |        |        |        |
| 2011 | 1.15% | 5.21%  | 0.73%  |        |
| 2012 | 0.97% | 13.30% | −1.54% |        |
| 2013 | 7.81% | 9.86%  | 4.39%  |        |
| 2014 | 4.94% | 5.45%  | 4.41%  |        |
| 2015 | 3.91% | 10.12% | 9.17%  |        |
| 2016 | 3.77% | −6.16% | 1.52%  |        |
| 2017 | 2.95% | 12.38% | 7.25%  |        |
| 2018 | 5.30% | −0.37% | 0.86%  |        |
| 2019 | 6.13% | 8.24%  | 6.00%  |        |
| 2020 | 7.63% | −3.27% | 15.79% | −3.12% |
| 2021 | 5.80% | 1.43%  | 18.00% | 1.36%  |
| 2022 | 5.86% | 4.92%  | 0.43%  | 4.69%  |

**TABLE 3** Sources: EUROSTAT (2020c; 2020a; 2021b), NOI (2021), NSI (2020c) Author's calculations.

Then, supposing a return to normality in 2022, PEN's growth is pegged to GDP growth reaching 0.43%. Finally, social contributions are assumed to overgrow their upward trend (EUROSTAT 2021b) to supply for about 30% of the fall in salaries in 2020. As wages get more dynamic in 2021, the trend is smoothed somewhat downward, falling from 7.63% to 5.80%. Due to sustained GDP growth, the precrisis trend should resume in 2022 — when SOC's growth will be 5.86%. On this basis, PIT revenues in Bulgaria could have grown by as much as 22% in 2020. The economic recovery in 2021 and 2022 should allow for a further increase in revenues by about 17% and 6% in 2021 and 2022, respectively. By the end of the triennium, PIT revenues would be 51% higher than in 2019.

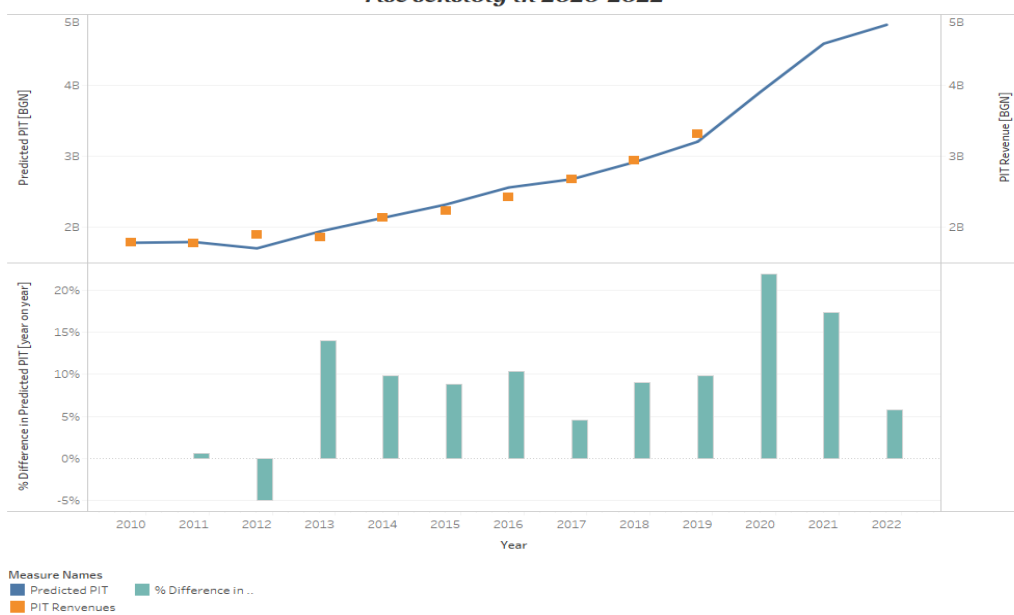

**Chart 2** Source: Author's calculations.

### 3.6 Total revenue forecasting error

The forecast-error statistics summarised in TABLE 4 (below) affirm the multivariate regression models' clear superiority for both PIT (panel A) and VAT (panel B). In comparison with official forecasts (MinFin na RB 2016), the multivariate regression models fare better than the 'baseline' ARIMA(1,1,1). The former has a smaller RMSE, median error (ME), median square error (MSE), mean absolute error (MAE), mean percentage error (MPE), mean absolute percentage error (MAPE) and standardised MAPE than either of the two alternatives. The DM and Theil's U tests confirm these results. Conversely, the ARIMA model underperforms official statics under all indicators but MSE and RMSE. Interestingly, official forecasts do not show the upwards bias alleged by some (Frankel 2011; Frankel and Schreger 2013).

Table 5: Forecast errors statistics show that the multivariate-regression model is the most precise of the alternatives presented here

| (5.A) | Regression | ARIMA(1,1,1) | Official data | (5.B) | Regression | ARIMA(1,1,1) | Official data |
|---|---|---|---|---|---|---|---|
| ME | −18.854 | −319.999 | −239.283 | ME | −10.389 | 711.099 | 649.652 |
| MSE | 7739.543 | 103010.139 | 105141.235 | MSE | 1075.783 | 551230.665 | 636294.484 |
| RMSE | 87.975 | 320.952 | 324.255 | RMSE | 32.799 | 742.449 | 797.681 |
| MAE | 71.670 | 319.999 | 287.667 | MAE | 29.892 | 711.099 | 649.652 |
| MPE | −0.011 | −0.119 | −0.101 | MPE | −0.001 | 0.075 | 0.065 |
| MAPE | 0.027 | 0.119 | 0.115 | MAPE | 0.003 | 0.075 | 0.065 |
| SMAPE | 0.027 | 0.112 | 0.107 | SMAPE | 0.003 | 0.078 | 0.068 |
| $U_1$ | 0.016 | 0.055 | 0.057 | $U_1$ | 0.002 | 0.041 | 0.044 |

TABLE 4 Source: Author's calculations.

That both VAT and PIT forecasts are systematically more reliable than both sophisticated baseline estimates and official data may lead to two observations. First, the common assumption as regards GDP growth may hold true almost surely in the sign and probably in the order of magnitude. Second, and more importantly, incorporating a few more variables are likely to improve forecasts' accuracy despite high uncertainty where reasonable estimates for the former can be inductively established.

## 4. Conclusions

### 4.1 Discussion of actual and forecasted tax revenue

CHART 1 (on page 22) and CHART 2 (on page 23) shows the actual and forecasted VAT and PIT revenues for the period 2010–2022. The same data are shown in TABLE 6 (below), where some trends are clearly evidenced.

Table 6: Results of the multivariate-regression estimate for PIT (6.A) and VAT (6.B)

| (6.A) | Predicted PIT | Residuals | Δ Predicted PIT | Δ SOC | Δ PEN | (6.B) | Predicted VAT | Residuals |
|---|---|---|---|---|---|---|---|---|
| 2010 | 1.78B | −192,088,057 | | | | 2010 | 6.4B | −52.5M |
| 2011 | 1.8B | −1,935,716 | 10.65M | 101.56M | 55.85M | 2011 | 6.6B | 4.2M |
| 2012 | 1.71B | 21,089,318 | −89.25M | 87.02M | −119.55M | 2012 | 7.4B | 1.8M |
| 2013 | 1.94B | 85,976,532 | 238.56M | 706.38M | 334.33M | 2013 | 7.6B | −9.8M |
| 2014 | 2.14B | −7,312,450 | 191.21M | 481.33M | 350.81M | 2014 | 7.5B | 4.4M |
| 2015 | 2.32B | 88,225,181 | 187.48M | 400.16M | 761.98M | 2015 | 8B | 36.2M |
| 2016 | 2.56B | 134,736,768 | 240.2M | 401.03M | 138.3M | 2016 | 8.7B | 26.5M |
| 2017 | 2.68B | 3,347,237 | 116.26M | 325.15M | 667.88M | 2017 | 9.2B | 38M |
| 2018 | 2.92B | −21,121,157 | 242.19M | 601.72M | 84.57M | 2018 | 10B | −5.2M |
| 2019 | 3.21B | −110,917,657 | 286.69M | 733.83M | 597.5M | 2019 | 11B | −43.5M |
| 2020 | 3.91B | | 702.55M | 968.54M | 1.67B | 2020 | 9.1B | |
| 2021 | 4.59B | | 677.69M | 793.17M | 2.2B | 2021 | 9.6B | |
| 2022 | 4.85B | | 265.27M | 847.63M | 61.41M | 2022 | 10.2B | |

TABLE 5 Sources: Official data of the Bulgarian government, Author's calculations

For a start, that PIT revenues could increase is not necessarily good news for anyone who cares about balancing the budget. As a matter of fact, the estimated growth in social transfers and pensions would vastly offset these gains and even cause a €2.5bln hole on the expenditure side. As observed above in relation to falling VAT revenues, the prospect of Bulgaria's debt and deficit skyrocketing in the next years is highly likely. True, Bulgaria's reputation as a 'fiscally responsible' country (Petkov 2014), its still small stock of liabilities (Hsing 2020) and the currency board make short-term deficits sustainable. However, if the debt were to balloon and – as it is expected – the internationally-low interest rate to burst (Amadeo 2021; Domm 2021; Mackenzie 2021), the risk of a debt-induced currency crisis could loom large again. Moreover, decreasing tax revenues may cause either a rate hike. Given the Bulgarian tax regime's marked regressivity (due to the adoption of flat-rate taxation across the board — Cf. Gaddy and Gale 2005; ECB 2007; OECD 2020), such a course of action may prejudice poorer households' capability of recovering from the crisis.

## CONCLUSION

This study allows to compare two different time series models' effectiveness and hint at multivariate regression as a better tool for forecasting during the pandemic for its greater flexibility. Moreover, the estimates shown above provide a perspective look at the country's future — surely a sketchy one, but troubling nonetheless. Some of these data actually do nothing but offering insights into the extent to which the pandemic has made facing already-pressing structural issues unavoidable for any political force. That the multivariate-regression models offer better forecasts than the government's official sources demonstrates the inaptness of the tools at Bulgarian policymakers' disposal in such a delicate juncture.

### Limitations of the study

The undertaken study has a few limitations. For instance, because of the currency board, the case of Bulgaria offers the advantage of ignoring exogenous shocks that may passthrough via the de- or appreciation of the exchange rate. Thus, future research studies may incorporate monetary factors in their estimates.

Furthermore, it was not possible to follow many of the IMF FAD's recommendations due to a lack of data. Most importantly, when it comes to VAT revenues it was impossible to adhere to the standard of a sectorial estimate of the tax base because neither the NSI's and EUROSTAT's adherence to NASE criteria make it difficult to identify concrete clusters of activities (e.g., tourism). The lack of high-frequency datasets is also an obstacle in attempting to consider other macroeconomic indicators such as inflation and unemployment. This hinderance was partly bypassed by diverting the analysis to correlated variables (i.e., employment income as a proxy for un- and under-employment). Yet, further studies may endeavour to explore this area.

## REFERENCES

[1] Alessi, Lucia, Eric Ghysels, Luca Onorante, Richard Peach, and Simon Potter. 2014. 'Central Bank Macroeconomic Forecasting During the Global Financial Crisis: The European Central Bank and Federal Reserve Bank of New York Experiences'. *Journal of Business & Economic Statistics* 32 (4): 483–500. https://doi.org/10.1080/07350015.2014.959124.

[2] Amadeo, Kimberly. 2021. 'When Will Interest Rates Go Up?' The Balance. 23 February 2021.

[3] Arminger, Gerhard, Clifford C Clogg, and Michael E Sobel. 1995. *Handbook of Statistical Modeling for the Social and Behavioral Sciences*. Boston, MA: Springer US : Imprint : Springer. https://doi.org/10.1007/978-1-4899-1292-3.

[4] Barrero, Jose Maria, Nicholas Bloom, and Steven J. Davis. 2020. 'COVID-19 Is Also a Reallocation Shock'. w27137. National Bureau of Economic Research.

[5] Blitz's Editors. 2021. 'ВМРО иска средна пенсия 700 лв. [IMRO wants an average pension of BGN 700]'. Blitz.bg. 4 March 2021.

[6] Blustein, David L., Ryan Duffy, Joaquim A. Ferreira, Valerie Cohen-Scali, Rachel Gali Cinamon, and Blake A. Allan. 2020. 'Unemployment in the Time of COVID-19: A Research Agenda'. *Journal of Vocational Behavior* 119 (June): 103436. https://doi.org/10.1016/j.jvb.2020.103436.

[7] Box, George E. P., Gwilym M. Jenkins, Gregory C. Reinsel, and Greta M. Ljung. 2016. *Time Series Analysis: Forecasting and Control*. 5th ed. Wiley Series in Probabilit and Statistics. Hoboken, New Jersey: John Wiley & Sons, Inc.

[8] Brinca, Pedro, João B. Duarte, and Miguel Faria-e-Castro. 2020. 'Measuring Sectoral Supply and Demand Shocks During COVID-19'. SSRN Scholarly Paper ID 3601938. Rochester, NY: Social Science Research Network.

[9] Brubakk, Leif, and Tommy Sveen. 2009. 'NEMO – a New Macro Model for Forecasting and Monetary Policy Analysis'. *Norges Bank - Economic Bulletin* 11 (July): 9.

[10] Carroll, Christopher D., and Miles S. Kimball. 1996. 'On the Concavity of the Consumption Function'. *Econometrica* 64 (4): 981–92.

[11] Carroll, Christopher D., Jiri Slacalek, Kiichi Tokuoka, and Matthew N. White. 2017. 'The Distribution of Wealth and the Marginal Propensity to Consume: The Distribution of Wealth'. *Quantitative Economics* 8 (3): 977–1020.

[12] Chakraborty, Shiney. 2020. 'COVID-19 and Women Informal Sector Workers in India'. *Economic and Political Weekly*, 5 June 2020.

[13] Darik news' Editors. 2020. 'Индексират Всички Пенсии с Между 5 и 10% От Юли Догодина [Indexing of All Pensions by between 5 and 10% from July next Year]'. DarikNews.Bg. 32 2020.

[14] Dellas, Harris, Dimitris Malliaropulos, Dimitris Papageorgiou, and Evangelia Vourvachaki. 2017.

'Fiscal policy with an informal sector'. Working Paper 235. Working Paper. Athens: Bank of Greece.
[15] Demertzis, Maria. 2021. 'Ανάκαμψη σχήματος K και δημοσιονομική πολιτική [K-shaped recovery and fiscal policy]'. *Kathimerini*, 2 March 2021, sec. Money Review.
[16] Demertzis, Maria, and Marta Dominguez-Jimenez. 2020. 'Monetary Policy in the Time of COVID-19, or How Uncertainty Is Here to Stay'. Policy paper. Monetary Dialogue Papers. Luxembourg: Committee on Economic and Monetary Affairs, Policy Department for Economic, Scientific and Quality of Life Policies, European Parliament. https://www.europarl.europa.eu/cmsdata/214970/02.BRUEGEL_final.pdf.
[17] Demertzis, Maria, and Nicola Viegi. 2008. 'Inflation Targets as Focal Points'. *International Journal of Central Banking* 4 (1): 55–87.
[18] ———. 2020. Steering the boat towards an unknown destination Interview by Giuseppe Porcaro. Podcast.
[19] Diacon, Paula-Elena, and Liviu-George Maha. 2015. 'The Relationship between Income, Consumption and GDP: A Time Series, Cross-Country Analysis'. *Procedia Economics and Finance*, 2nd GLOBAL CONFERENCE on BUSINESS, ECONOMICS, MANAGEMENT and TOURISM, 23 (January): 1535–43. https://doi.org/10.1016/S2212-5671(15)00374-3.
[20] Dickey, David A., and Wayne A. Fuller. 1979. 'Distribution of the Estimators for Autoregressive Time Series with a Unit Root'. *Journal of the American Statistical Association* 74 (366a): 427–31.
[21] ———. 1981. 'Likelihood Ratio Statistics for Autoregressive Time Series with a Unit Root'. *Econometrica* 49 (4): 1057–72. https://doi.org/10.2307/1912517.
[22] Diebold, Francis X, and Robert S Mariano. 2002. 'Comparing Predictive Accuracy'. *Journal of Business & Economic Statistics* 20 (1): 134–44.
[23] Domm, Patti. 2021. 'Rising Interest Rates May Continue to Test the Stock Market in the Week Ahead'. CNBC. 26 February 2021.
[24] DoT Australia. 2008. 'Forecasting Accuracy of the Act Budget Estimates'. Department of Treasury and Finance, Government of Australia. http://www.treasury.act.gov.au/documents/Forecasting%20Accuracy%20-%20ACT%20Budget.pdf.
[25] Draganov, Nikolai. 2020. 'Рестрикции – добре, но къде са мерките за засегнатите? [Restrictions - well, but where are the measures for those affected?]'. Barikada. 15 March 2020. https://baricada.org/2020/03/15/sb-merki/.
[26] ECB. 2007. 'Flat Taxes in Central and Eastern Europe'. In *Monthly Bulletin - September 2007*, 81–83. Munich (Germany): European Central Bank.
[27] ———. 2020. 'Eurosystem Staff Macroeconomic Projections for the Euro Area, June 2020'. Eurosystem Staff Macroeconomic Projection for the Euro Area. Frankfurt am Main: European Central Bank.
[28] Estupinan, Xavier, and Mohit Sharma. 2020. 'Job and Wage Losses in Informal Sector Due to the COVID-19 Lockdown Measures in India'. SSRN Scholarly Paper ID 3680379. Rochester, NY: Social Science Research Network. https://doi.org/10.2139/ssrn.3680379.
[29] EU Commission. 2020. 'European Economy Forecast 2020, Autumn.' Forecast. European Economic Forecast.
[30] EUROSTAT. 2008. *NACE Rev. 2*. Luxembourg: Office for Official Publications of the European Communities.
[31] ———. 2017. 'Income Inequality and Poverty Indicators'. EUROSTAT - Experimental Statistics. 3 October 2017.
[32] ———. 2020a. 'Impact of COVID-19 on Employment Income - Advanced Estimates'. EUROSTAT - Experimental Statistics. October 2020.
[33] ———. 2020b. 'Structure of Household Population by Activity Status of the Reference Person'. EUROSTAT Data. 7 December 2020.
[34] ———. 2020c. 'COVID-19 Impact on Employment Income'. EUROSTAT - Experimental Statistics. 10 December 2020.
[35] ———. 2021a. 'GDP per Capita in PPS'. EUROSTAT Data. 15 February 2021.
[36] ———. 2021b. 'Social Benefits (Other than Social Transfers in Kind) Paid by General Government'. EUROSTAT Data. 15 February 2021.
[37] Fisher, Jonathan D., David S. Johnson, Timothy M. Smeeding, and Jeffrey P. Thompson. 2020. 'Estimating the Marginal Propensity to Consume Using the Distributions of Income, Consumption, and Wealth'. *Journal of Macroeconomics* 65 ([103218]): 1–14. doi: 10.1016/j.jmacro.2020.103218.
[38] Frankel, Jeffrey. 2011. 'Over-Optimism in Forecasts by Official Budget Agencies and Its Implications'. *Oxford Review of Economic Policy* 27 (4): 536–62. https://doi.org/10.1093/oxrep/grr025.
[39] Frankel, Jeffrey, and Jesse Schreger. 2013. 'Over-Optimistic Official Forecasts and Fiscal Rules in the Eurozone'. *Review of World Economics* 149 (2): 247–72. https://doi.org/10.1007/s10290-013-0150-9.
[40] Gaddy, Clifford G., and William G. Gale. 2005. 'Demythologizing the Russian Flat Tax'. *Tax Notes International*, 14 March 2005.
[41] Granger, C. W. J. 1969. 'Investigating Causal Relations by Econometric Models and Cross-Spectral Methods'. *Econometrica* 37 (3): 424.
[42] Guerrieri, Veronica, Guido Lorenzoni, Ludwig Straub, and Iván Werning. 2020. 'Macroeconomic Implications of COVID-19: Can Negative Supply Shocks Cause Demand Shortages?' w26918. National Bureau of Economic
[43] Harvey, A. C., and P. H. J. Todd. 1983. 'Forecasting Economic Time Series with Structural and Box-Jenkins Models: A Case Study'. *Journal of Business & Economic Statistics* 1 (4): 299–307. https://doi.org/10.2307/1391661.
[44] Heckelman, Jac C. 2000. 'Economic Freedom and Economic Growth: A Short-Run Causal Investigation'. *Journal of Applied Economics* 3 (1): 71–91.

[45] Hsing, Yu. 2020. 'On the Relationship between Economic Growth and Government Debt for Bulgaria. Test of the Reinhart-Rogoff Hypothesis'. *Theoretical and Applied Economics* 27 (4): 187–94. http://store.ectap.ro/articole/1502.pdf.
[46] ILO. 2020. 'Young Workers Will Be Hit Hard by COVID-19's Economic Fallout'. *Work In Progress - International Labour Organization* (blog). 15 April 2020.
[47] IMF FAD. 2020. 'Challenges in Forecasting Tax Revenue'. Research note. Special Series on Fiscal Policies to Respond to COVID-19. Washington, D.C.: International Monetary Fund.
[48] Ivanov, Dobrin. 2021. Защо АИКБ предлага работниците да плащат 100% от осигуровките си? [Why does BICA offer the workers to pay 100% of their insurances?] Interview by Hristo Nikolov. Bloomberg TV - Bulgaria. https://www.bloombergtv.bg/a/16-biznes-start/89473-aikb-nedoverieto-v-darzhavata-i-niskite-dohodi-stimulirat-nereglamentiranata-zaetost.
[49] Karagyozova-Markova, Kristina, Georgi Deyanov, and Viktor Iliev. 2013. *Fiscal Policy and Economic Growth in Bulgaria*. Discussion Papers / Bulgarian National Bank 90. Sofia.
[50] Kato, Junko. 2003. *Regressive Taxation and the Welfare State: Path Dependence and Policy Diffusion*. Cambridge University Press.
[51] Keen, Mr Michael. 2013. *The Anatomy of the VAT*. Washington D.C.: International Monetary Fund.
[52] Kirchgässner, Gebhard, Jürgen Wolters, and Uwe Hassler. 2013. *Introduction to Modern Time Series Analysis*. Springer Texts in Business and Economics. Berlin, Heidelberg: Springer Berlin Heidelberg.
[53] Kolasa, Aleksandra. 2012. 'Life Cycle Income and Consumption Patterns in Transition'. SSRN Scholarly Paper ID 2210348. Rochester, NY: Social Science Research Network. https://doi.org/10.2139/ssrn.2210348.
[54] Legeida, Nina, and Dimitry Sologoub. 2003. 'Modeling Value Added Tax (VAT) Revenues in a Transition Economy: Case of Ukraine'. Working Paper 23. Working Paper. Kiev: Institute for Economic Research and Policy Consulting. http://www.ier.com.ua/files/publications/WP/2003/WP22_eng.pdf.
[55] Lex.bg's Editors. 2020. 'С 9% ДДС ще са ресторанти, хотели, книги, детски храни и памперси [The 9% VAT will be restaurants, hotels, books, children's food and diapers]'. Lex.bg News. 10 June 2020.
[56] Mackenzie, Michael. 2021. 'Should Equity Investors Worry about Rising Interest Rates?' The Financial Times. 20 February 2021.
[57] Mediapool.bg's Editors. 2020. 'Сивата икономика се е свила до 21%, но липсата на реформи забавя темпа [The informal economy has shrunk to 21%, but the lack of reforms is slowing down ]'. Mediapool.bg. 15 December 2020.
[58] Milcheva, Emilia. 2020. 'Коронавирус: какви мерки предприема българското правителство [Coronavirus: what measures the Bulgarian government is taking]'. Deutsche Welle. 15 March 2020.
[59] MinFin na RB. 2016. 'Средносрочна бюджетна прогноза за периода 2017 - 2019 г. [Mid-term budget forecast for the period 2017 - 2019]'. Ministerstvo na Finansiite na Republika Bŭlgaria.
[60] Morozumi, Atsuyosh, and Santiago Acosta Ormaechea. 2019. 'The Value Added Tax and Growth: Design Matters'. Working Paper 96. IMF Working Papers. Washington D.C.: International Monetary Fund.
[61] Narula, Rajneesh. 2020. 'Policy Opportunities and Challenges from the COVID-19 Pandemic for Economies with Large Informal Sectors'. *Journal of International Business Policy* 3 (3): 302–10.
[62] Nau, Robert. 2014. 'Mathematical Structure of ARIMA Models'. Fuqua School of Business, Duke University.
[63] Nikolova, Glorina. 2020. 'Обмислят Две Увеличения На Пенсиите През 2021 г. [Two Pension Increases under Consideration in 2021]'. БТВ Новините. 8 October 2020.
[64] NOI. 2020a. 'Добавки Към Пенсии [Pension Supplements]'. Natsionalen Osiguritelen Institut. 2020.
[65] ———. 2020b. 'Пенсиите и Добавките От 50 Лева Към Тях Ще Се Изплащат Между 7 и 21 Декември [Pensions and Supplements of BGN 50 to Them Will Be Paid between December 7 and 21]'. Natsionalen Osiguritelen Institut. 30 November 2020.
[66] ———. 2021. 'Справка За Среден Основен Размер На Водещата Пенсия [Information on the Average Basic Amount of the Leading Pension]'. Natsionalen Osiguritelen Institut. 15 March 2021.
[67] NSI. 2020a. 'Население и демографски процеси през 2019 година [Population and demographic processes in 2019]'. Natsionalen Statisticheski Institut.
[68] ———. 2020b. 'БВП - Производствен Метод - Национално Ниво [GDP - Production Method - National Level]'. Natsionanel Statisticheski Institut. 19 October 2020.
[69] ———. 2020c. 'Общ Доход На Домакинствата [Total Household Income]'. Natsionanel Statisticheski Institut. 22 October 2020.
[70] NS-RB. 2020. *Закон за Българската народна банка [Law on the Bulgarian National Bank]*.
[71] OECD. 2020. 'Tax on Personal Income'. OECD Data. 2020.
[72] Petkov, Vasil. 2014. 'Advantages and Disadvantages of Fiscal Discipline in Bulgaria in Times of Crisis'. *Contemporary Economics* 8 (1): 47–56.
[73] Rosser, J. Barkley, Marina V. Rosser, and Ehsan Ahmed. 2000. 'Income Inequality and the Informal Economy in Transition Economies'. *Journal of Comparative Economics* 28 (1): 156–71.
[74] Streimikiene, Dalia, Rizwan Raheem Ahmed, Jolita Vveinhardt, Saghir Pervaiz Ghauri, and Sarwar Zahid. 2018. 'Forecasting Tax Revenues Using Time Series Techniques – a Case of Pakistan'. *Economic Research-Ekonomska Istraživanja* 31 (1): 722–54.

[75] Tamaoka, Masayuki. 1994. 'The Regressivity of a Value Added Tax: Tax Credit Method and Subtraction Method - A Japanese Case'. *Fiscal Studies* 15 (2): 57–73. https://doi.org/10.1111/j.1475-5890.1994.tb00197.x.
[76] Telarico, Fabio Ashtar. 2021. 'The US's Schizophrenic Recovery: Banks' Earnings on the Rise as the Government Bails out Families'. Global Risk Insights. 12 March 2021.
[77] Tiao, George C., and George EP Box. 1981. 'Modeling Multiple Time Series with Applications'. *Journal of the American Statistical Association*.
[78] Tosun, Mehmet Serkan, and Sohrab Abizadeh. 2005. 'Economic Growth and Tax Components: An Analysis of Tax Changes in OECD'. *Applied Economics* 37 (19): 2251–63. https://doi.org/10.1080/00036840500293813.
[79] Toteva, Paolina. 2021. 'Работодателите: Недекларираната заетост е най-масовата сива практика [Employers: Undeclared employment is the most common gray practice]'. *Flashnews* (blog). 22 February 2021.
[80] Trud's Editors. 2020. 'ВМРО Приема Индексиране На Пенсиите [IMRO Accepts Indexation of Pensions]'. Trud.Bg. 15 September 2020.

**SUMMARY**

Tax analysis and forecasting of revenues is of paramount importance to ensure the viability and sustainability of fiscal policy. However, the measures taken to contain the spread of the recent pandemic pose an unprecedented challenge. This paper proposes a model to forecast tax revenues in Bulgaria for the fiscal years 2020–2022 built in accordance with the International Monetary Fund's recommendations. This study allows to compare two different time series models' effectiveness. The outcomes hint at multivariate regression as a better tool for forecasting during the pandemic for its greater flexibility. Moreover, the estimates shown above provide a perspective look at the country's future — surely a sketchy one, but troubling nonetheless. Some of these data actually do nothing but offer insights into the extent to which the pandemic has made facing already-pressing structural issues unavoidable for any political force. That the multivariate regressionmultivariate regression models offer better forecasts than the government's official sources demonstrates the inaptness of the tools at Bulgarian policymakers' disposal in such a delicate juncture.